\documentclass[floatfix]{emulateapj} %approximates print-look
\shorttitle{Jets as diagnostics of the CSM}
\shortauthors{Schure et al.}

\begin{document}

\title{Jets as diagnostics of the circumstellar medium and the explosion energetics of supernovae: the case of Cas A}
\author{K.M. Schure\altaffilmark{1}, J. Vink \altaffilmark{1}, G. Garc\'ia-Segura \altaffilmark{2} and A. Achterberg\altaffilmark{1}}
\altaffiltext{1}{Astronomical Institute, University of Utrecht,
             Postbus 80000, NL-3508 TA Utrecht; K.M.Schure@phys.uu.nl, J.Vink@uu.nl, A.Achterberg@uu.nl}
\altaffiltext{2}{Instituto de Astronom\'ia, Universidad Nacional Aut\'onoma de M\'exico, 22800 Ensenada, Mexico; GGS@astrosen.unam.mx}

\begin{abstract}
We present hydrodynamical models for the Cassiopeia A (Cas A) supernova remnant and its observed jet / counter-jet system. We include the evolution of the progenitor's circumstellar medium, which is shaped by a slow red supergiant wind that is followed by a fast Wolf-Rayet (WR) wind.  

The main parameters of the simulations are the duration of the WR phase and the jet energy. We find that the jet is destroyed if the WR phase is sufficiently long and a massive circumstellar shell has formed.  We therefore conclude that the WR phase must have been short (a few thousand yr), if present at all. Since the actual jet length of Cas A is not known we derive a lower limit for the jet energy, which is $\sim 10^{48}$~erg. We discuss the implications for the progenitor of Cas A and the nature of its explosion. 
\end{abstract}


\section{Introduction}

Over the last decade evidence has emerged that suggests that at least some core-collapse supernovae are intrinsically non-spherically symmetric explosions. The evidence is strongest for supernovae of stars that have lost most of their outer (hydrogen-rich) envelopes, i.e., the type Ib/c supernovae \citep{2001Wangetal}. 
For those explosions, the inner layers are exposed early on, and asymmetries in the core more easily survive the interactions with the outer layers. This implies that departures from spherical symmetry originate from deep inside the explosion.

A better understanding of the explosion geometries is needed to provide further insights into what powers core-collapse  supernovae. In the canonical explosion model the explosion is driven by deposition of neutrino energy into the region just outside the proto-neutron star. However, up to now, computer simulations of this core collapse do not self-consistently predict supernova explosions  \citep{2007Jankaetal}. In those simulations the role of magnetic fields and stellar rotation is usually neglected. According to \citet{1999Khokhlovetal} and \citet{2000Wheeleretal}, magnetic fields and rotation may play a crucial role in the explosion mechanism, and may lead to bipolar explosions. Additionally, other explosion mechanisms that are based on acoustic and hydrodynamic instabilities can result in asymmetric, albeit not necessarily bipolar explosions \citep[e.g.][]{2007BlondinShaw,2007Burrowsetal}. 
However, if one considers the most energetic supernova explosions; those associated with the long duration gamma-ray bursts (LGRBs), it is very likely that these explosions are truly bipolar. The associated supernovae are of type Ic, see \citet{2006DellaValle} for a review. 

The engines that drive the explosions associated with LGRBs may or may not be related to those of ``normal'' core-collapse supernovae. In the collapsar model, LGRBs are the result of black hole formation \citep{1999MacFadyenWoosley} and thus have a distinctly different engine from that producing normal supernovae. Alternative models that consider LGRBs to be powered by highly magnetic, rapidly rotating, neutron stars \citep[e.g.][]{2004Thompsonetal} or by trans-relativistic blast waves in supernovae \citep{2001Tanetal}, allow for a continuum of bipolarity and explosion energies. In those cases, the amount of rotation and magnetic field strength, and the line of sight, determine whether we observe a ``normal'' supernova or one associated with a LGRB. 
This study aims to shed some light on the intermediate case of a supernova that shows distinct bipolarity, but is energetically in the range of regular supernovae and does not have relativistic ejecta. We try to provide some insight in the requirements on the energy in the asymmetric part of the supernova, i.e. it's ``jets'', and on the type of progenitor that was responsible for the circumstellar medium (CSM) at time of explosion.

%-------------------------------------------------------------
%\clearpage
  \begin{figure}[!htbp]
%  \begin{minipage}{\textwidth}
   \centering
   \plotone{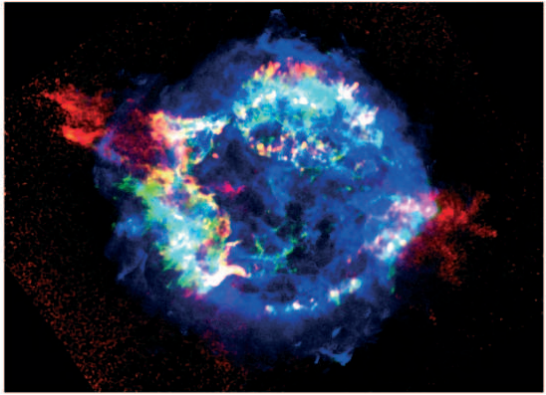}

      \caption[ ] {Three color image showing the location of the jets \citep[red, see][]{2004Vink,2004Hwangetal} with respect to the bright X-ray shell of ejecta (green, X-ray Si~\textsc{XIII} emission) and radio synchrotron emission (VLA archival data). The jet image is obtained by taking the ratio of Si~{\sc XIII} over Mg~{\sc XI} X-ray line emission. (Public domain image based on the 1~Ms Chandra observation of  Cas A \citep{2004Hwangetal} [http://www.astro.uu.nl/$\sim$vinkj/casa\_jet\_si\_radio.jpg]).
      \label{fig:casa_SiMg}}
%      \end{minipage}
   \end{figure}
%\clearpage
%-------------------------------------------------------------

Two likely examples of bipolar supernovae are known in the local neighborhood: SN1987A \citep{2002Wangetal} and the supernova remnant (SNR) Cassiopeia A, the subject of this paper. The bipolarity of Cas A has only recently been established from optical \citep{2001Fesen,2006Fesenetal}, X-ray \citep{2004Vink, 2004Hwangetal,2006Lamingetal} and infrared \citep{2004Hinesetal} observations. These observations show that, apart from the long known ``jet'' region in the northeast, a somewhat less prominent protrusion is located in the southwest (Fig.~\ref{fig:casa_SiMg}). In Figure 1 we show in red the image of the jet as shown in \citet{2004Hwangetal}. In order to show the jet in the context of the overall emission it is combined with images in silicon (green) and radio (blue). For details and a discussion on the jet and its abundances we refer to \citet{2004Vink}, \citet{2004Hwangetal}, and \citet{2006Fesenetal}. 

The jets extend out to a radius of at least 3.8~pc, for a distance of 3.4~kpc \citep{1995Reedetal}. The reason to believe that these jets are the results of a bipolar explosion, rather than being caused by a bipolar structure in the CSM \citep{1996Blondinetal},  is the distinct elemental abundance patterns in both jet regions, with the jet material coming from deeper layers inside the star. The X-ray and optical data indicate that the jet material is rich in oxygen burning products (Si, S, Ar, Ca), while it lacks carbon- and neon-burning products (O, Ne, Mg). This is the reason why the jet, of which the emissivity is weak compared to the rest of Cas A, stands out by taking the ratio of the Si/Mg line emission.

Interestingly, the supernova that caused Cas A seems to have some shared characteristics with the supernovae associated with LGRBs and X-ray flashes. Cas A's progenitor probably had lost most of its hydrogen envelope, given the lack of hydrogen rich, optically identified, ejecta. Although we are not claiming that Cas A was an LGRB or even an X-ray flash, the two non-relativistic jets suggest that it may be related and there may be a continuum of bipolarity in the explosion of supernovae, thus providing a possible link between LGRBs and normal supernovae.
The  total mass of shocked ejecta is 2-4~M$_\odot$  \citep{1996Vinketal}, and the explosion energy is about a factor of two more than the canonical explosion energy of $10^{51}$~erg  \citep{2003HwangLaming}. The total oxygen ejecta mass of 1-2~M$_\odot$ suggests a main sequence mass of $18-22$~M$_\odot$ \citep{2004Vink}. These properties are reminiscent of the parameters derived for SN2006aj, the supernova associated with the X-ray flash XRF 060218  \citep{2006Mazzalietal}, and similar to SN2003jd, the one suggested to relate to a LGRB\citep{2007Valentietal}.

The large amount of swept up mass in Cas A and the dynamic properties of the blast wave suggest that the blast wave is currently moving through the high-density red supergiant (RSG) wind \citep{2003ChevalierOishi, 2004Vink}. However, the lack of H-rich ejecta suggests that Cas A exploded as a Wolf-Rayet (WR) star. Moreover, the presence of slow moving N-rich knots has been explained as originating from the hydrodynamical instabilities between the fast WR wind and the dense, slow moving, RSG wind \citep{1996GGSetal_II}. 

In this paper we present hydrodynamical simulations of the jets in the context of the progenitor's mass loss history, which we take to be a RSG phase, possibly followed by a WR phase.
There are two main reasons for pursuing this problem. First of all, the energetics of the jets can be better estimated using a realistic mass loss history in the hydro-simulations. Secondly, the survival of the jets depends strongly on the mass-loss history of the progenitor. Therefore, the jets in Cas A can be used as a diagnostic on both the properties of the bipolar explosion, and on the progenitor-shaped circumstellar medium (CSM) at the time of explosion.

%__________________________________________________________________

\section{Methods}
\label{sec:methods}

The simulations of the supernova explosion and the progenitor winds are done using the code {\tt ZEUS-3D v3.4} \citep{1996Clarke}, an extended version of the {\tt ZEUS-2D} code by \citet{1992StoneNorman}. This code solves the ideal non-relativistic fluid equations in three dimensions by finite differencing on an Eulerian mesh. Von-Neumann Richtmyer artificial viscosity is used to deal with shocks. Energy losses by radiative cooling are calculated according to the cooling curve as described by \citet{1981MacDonaldBailey}. Photo-ionization is used as implemented by \citet{1999GarciaSeguraetal}, where matter within the Str\"omgren radius is fully ionized, whereas all the rest is considered neutral.

In our case the code is set up in spherical ($r \theta \phi$) coordinates with the assumption of symmetry around the polar $\phi$-axis. Since the supernova remnant does not reach beyond the radius where the RSG wind meets the main sequence bubble, the initial grid is filled with a RSG wind. This part of the simulation is done in 1D in order to save on computational time. Once the RSG wind occupies the whole grid, the resulting CSM is transferred onto a 2D grid with 900 cells in the radial direction and 225 cells in the angular direction. The equidistant grid cells are distributed over a radial extent of 6~pc and an angular extent of $\pi/4$ measured from the pole, meaning an effective resolution of $2.1 \times 10^{16}$~cm by $0.2^\circ$. The applied resolution appears to be adequate for our purposes: doubling the resolution did not affect the hydrodynamical evolution of the structure and development of the instabilities.

Into this grid, we release a WR wind that lasts for a period of $5.5 \times 10^{11}$~s ($\sim 17,000$~yr). 
During the simulation of this evolutionary stage of the supernova progenitor, every $5.5 \times 10^9$~s a snapshot is taken of the CSM. These snapshots, representing the CSM for different durations of the WR phase of the progenitor, serve as the initial conditions in which we subsequently add the supernova ejecta and follow the evolution of the SNR. In other words, the age of the WR phase at the time of the explosion is a variable in the different simulations of the SNR, depending on the snapshot we take of the progenitor evolution to serve as the initial condition. 
The maximum duration of the Wolf-Rayet phase is restricted by the high amount of mass in the Cas A remnant and the high density the blast wave is currently running into, indicating that the swept-up shell, hereafter called WR shell, is within the border of the remnant (i.e. 2.5~pc). 

The parameters for the RSG and WR winds are taken from van Veelen \& Langer (in prep. 2008). The properties of the RSG wind are determined by the amount of mass lost by the progenitor star and the typical RSG life time as in the model by \citet{2004HirschiMeynetMaeder}. This gives a mass loss rate and terminal wind velocity of $\dot M=1.54 \times 10^{-5}$~M$_\odot$yr$^{-1}$ and $v_{\rm RSG}=4.7$~km~s$^{-1}$. The temperature of the RSG wind is set to the typical value for this type of wind: $T=10^3$~K. The properties of the WR wind: $\dot M=9.7 \times 10^{-6}$~M$_\odot$~yr$^{-1}$, $v=1.7 \times 10^3$~km~s$^{-1}$, and $T=10^4$~K, were calculated in accordance with \citet{2000NugisLamers}. Note that qualitatively our results do not depend on the exact parameter chosen here, as will be discussed in section~\ref{sec:SDJS}.   Observational constraints for the shocks are taken from several authors \citep{2003DeLaneyRudnick,1998Vinketal,2004Morseetal,1998Koraleskyetal} to be the following: $v_{\rm forward} \approx 5000$~km~s$^{-1}$, $r_{\rm forward} \approx 2.5$~pc, and the reverse shock to: $v_{\rm reverse} \approx 3000$~km~s$^{-1}$, $r_{\rm reverse} \approx 1.4$~pc. 

The explosion parameters are taken from \citet{2003HwangLaming} and \citet{2004Vink} to be $E_{\rm ej}=2 \times 10^{51}$, $M_{\rm ej}=2-4$~M$_\odot$, where we adopt an ejecta mass of $M_{\rm ej}=2.5$~M$_\odot$. The initial ejecta density profile covers 0.1~pc in radius, and consists of a constant-density core, with an envelope for which the density decreases as $\rho \propto r^{-9}$, which is the typical density profile of the ejecta in explosion models \citep[c.f.][]{1999TrueloveMcKee}. In order to match the observationally determined ejecta mass and energy, we iteratively determine the value for the density in the ejecta core and the velocity at which the core ends and the powerlaw envelope begins. For the parameters adopted in our simulations, this happens for a central density of the ejecta core of $1.1 \times 10^{-19}$g~cm$^{-3}$, and the powerlaw envelope begins where the velocity of the ejecta is $9.8 \times 10^8$cm~s$^{-1}$. The velocity linearly increases from zero at the core, to 15,000 km~s$^{-1}$ at the outer part of the envelope.

Since the details of the jet-forming mechanism are poorly understood, we try to explore the relevant parameter space in density and velocity by adjusting them in the jet region and thus locally enhancing the kinetic energy. Since it is not to be expected that the pre-supernova density will be much higher in the jet region, we focus on the velocity, although we do explore a factor of two in density decrease and enhancement. The shapes of the density and velocity profiles are expected to be the same as in the rest of the ejecta, since these are determined by the propagation of the explosion through the progenitor star \citep{1999MatznerMcKee}.

\section{Results}
\label{sec:results}

\begin{figure*}[!htbp]
   \centering
   \plotone{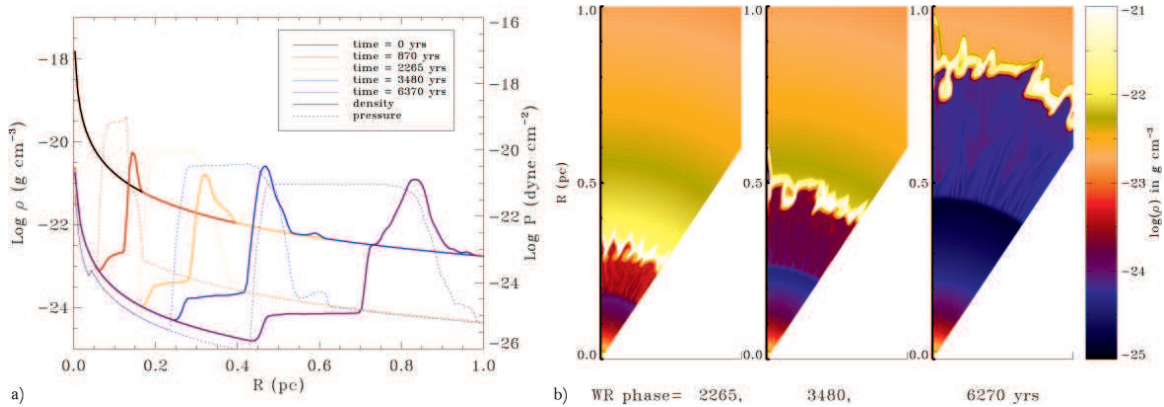}
      \caption{{\em a}) CSM density (solid lines) and pressure (dotted) due to the RSG and WR winds are shown, for different stages in the evolution.  The black solid line shows only the free-streaming RSG wind. For longer WR phases the WR shell broadens and the density contrast increases.  The shocks appear smooth due to averaging over the $\pi/4$-angle we used in the 2D simulation. 
      {\em b}) CSM density evolution is plotted in 2D, which exemplifies the irregularities in the shell due to the Vishniac thin shell instability. The three panels show the CSM for different durations of the WR phase and correspond to the last three profiles in {\em a}). The forward shock is marked by the sudden increase in density and corresponds to the outer boundary of the shell, at a radius of approximately $0.35$, $0.5$, and $0.9$~pc. The contact discontinuity corresponds to the sudden drop in density, and the inner boundary of the thin shell. The wind-termination shock is located at around $0.15$, $0.25$ and $0.45$ in the three panels respectively.} 
         \label{fig:WRshell}
   \end{figure*}

During the progenitor's evolution, the CSM is shaped by a RSG wind that may be followed by a WR wind. The WR wind sweeps up a shell, whose radius and mass depend on the duration of the WR phase.
In Figure~\ref{fig:WRshell} the result of a 2D simulation of this evolution is plotted in 1D and 2D. The swept-up shell is irregular due to hydrodynamic instabilities. Since both winds are steady, the shell moves with constant velocity and is not Rayleigh-Taylor unstable. However, because of the large velocity difference between the WR wind and the RSG wind, and because cooling is implemented in the code and effective in the shell, the shell becomes sufficiently thin to be susceptible to the Vishniac ``thin shell'' instability \citep{1983Vishniac}. This instability arises when the shell is thin enough and a compression ratio of over $\sim 21$ is reached, which is the case in our simulations. At late times in the WR shell evolution, the shell becomes more and more fragmented. However, this only occurs beyond the point we consider here, which is limited by the requirement that the SNR is currently running into the RSG wind and the shell therefore must be within a radius of 2.5~pc. For a more detailed description of the instabilities and when they occur, we refer to \citet{1995GGSMacLow,1996GGSetal_II}. We find that doubling the resolution in both $r$ and $\theta$, reaching an effective resolution of $1.0 \times 10^{16}$~cm by $0.1^\circ$, does not make a qualitative difference in the appearance of the instabilities and the overall dynamics. 

Figure~\ref{fig:WRshell}{\em a} shows an average over all angles, which smooths out the density and pressure jumps. The free streaming RSG and WR wind follow the density profiles $\left(\rho = \dot M/(4 \pi r^2 v_{\rm wind})\right)$ that depend on their respective mass loss rate and velocity. In the swept up shell the pressure is much higher than outside the shocked region. The shell consists, farthest from the star, of shocked, compressed RSG wind material. Further inward, the contact discontinuity, characterized by a jump in the density but with a constant pressure, separates the shocked RSG wind from the shocked WR wind.
Figure~\ref{fig:WRshell}{\em b}  shows the density of the CSM in 2D. The Vishniac thin shell instability causes irregularities in the swept up shell. The three panels in Figure~\ref{fig:WRshell}{\em b} correspond to the last three evolutionary stages that were plotted in Figure~\ref{fig:WRshell}{\em a}. Some accumulation of mass in the shell near the polar axis is due to the boundary conditions. This however does not qualitatively influence the outcome of the subsequent supernova simulation, as we tested with a simulation of the winds in the $r\phi$-plane. 

%-------------------------------------------------------------
%\clearpage
%\clearpage

The CSM that is created in this manner provides the background in which we implement the asymmetric supernova ejecta. The main parameters that we vary are the duration of the WR phase of our background and the energy in the jet. In the following part we focus on the results of different jet energies and use the background of a plain RSG wind.

\subsection{Jet Evolution in a RSG Wind}
\begin{figure}[!htbp]
   \centering
 \plotone{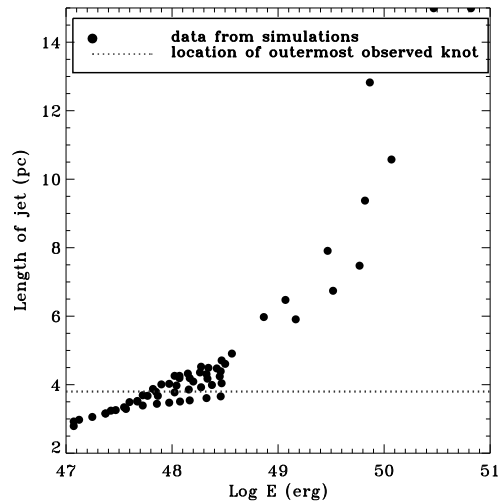}
      \caption{Data from simulations (black dots), representing the length of the jet for different jet energies. We have varied the initial opening angle, density and, mostly, the velocity, thus sometimes getting different jet lengths for similar energies. If the mass in the jet is relatively high, the jet protrudes further. The horizontal line indicates where the outermost observed ejecta are located. }
         \label{fig:length_energy}
   \end{figure}

With the explosion parameters as given in section~\ref{sec:methods}, we find for the equatorial forward and reverse shock the following properties:  $v_{\rm forward} = 5430$~km~s$^{-1}$, $r_{\rm forward} = 2.43$~pc, and $v_{\rm reverse} = 3160$~km~s$^{-1}$, $r_{\rm reverse} =1.63$~pc, for an age of the remnant of 330~yr. 
The jet is collimated by the high pressure cocoon that results from its bow shock. The initial opening angle is taken to be very small (half opening angle of about $1^\circ$). Variations of this opening angle have been tried and we find that as long as it is not too large ($< 4^\circ$), the half opening angle of the jet after 330 years is always about $10^\circ$ \citep[similar to what is observed in Cas A,][]{2006Lamingetal}.  In numerous low-resolution simulations we have explored the effect of different jet densities and velocities on the resulting jet length. These results are shown in Figure~\ref{fig:length_energy}. Differences in jet length for the same jet energy derive from the relative contribution of mass and velocity. For a given energy, a higher density gives rise to a longer jet. 

In order to see how the length of the jet scales with its energy, we have derived analytical solutions for both a cone-shaped jet with constant opening angle and for a cylindrically shaped jet (Appendix~\ref{app:lEjet}). These give the following relations between the length of a jet and its energy, for a cone:

\begin{eqnarray}
\label{eq:lE13}
R&=&1.15\left( \frac{E t^2}{ \theta_{\rm j}^2 A} \right)^{1/3},
\end{eqnarray}
where $\theta_{\rm j}$ is the opening angle, $E$ the jet energy, $R$ the length of the jet at time $t$, and $A$ is a density normalization constant.
For a cylinder the relation is:

\begin{eqnarray}
R =  \sqrt{\frac{E t^2 R_0}{C_\gamma A \pi r^2}} ,
\end{eqnarray}
with $r$ the radius of the cone and $C_\gamma = (4 \gamma)/((\gamma-1)(\gamma+1)^2) = 45/32$ for a polytropic index of $\gamma=5/3$.
Neither of these relations gives a satisfactory fit to the simulated jet length. Clearly, the approximations used are too simple to encompass all the relevant physics. For example, the assumptions are based on a point explosion and neglect the presence of a reverse shock, lateral spreading of the jet and interaction with the rest of the supernova, which are also important for the propagation of the jet. However, it gives us an estimate of how the propagation of the jet may scale with energy and the order of magnitude we can expect.

%-------------------------------------------------------------
%\clearpage
%\clearpage
%

In the ``jet''-region, optical knots have been observed out to a distance of 3.8~pc from the center of the SNR \citep{2001Fesen}. As mentioned by \citet{2006Lamingetal} this may not be the actual tip of the jet, which may be invisible due to cooling. The actual blast wave may be outside of the field of view of available X-ray observations. The same authors estimate the tip to be at a distance of 5.66~pc, with an equivalent isotropic energy for the jet of $2.3 \times 10^{52}$~erg. Our findings concur with these estimates. For a jet length of $3.8 - 6$~pc, our simulations show that the energy of the jet should be in between $6.0 \times 10^{47}$~erg and $5.0 \times 10^{48}$~erg, which is equivalent to $8.0 \times 10^{51}$~erg and $6.6 \times 10^{52}$~erg isotropic energy.

\subsection{The Evolution of a Jet in the Presence of a WR Shell}
\label{sec:jetinwrshell}
The presence of a WR shell severely alters the morphology of the SNR and limits the survival of a jet. From low-resolution simulations we find that for WR shells further out than $\sim 1$~pc the forward shock of the SNR extends beyond the observed 2.5~pc. We therefore focus our higher resolution simulations on the earliest stages of the WR phase. 

\begin{figure*}[!htbp]
   \centering
   \plotone{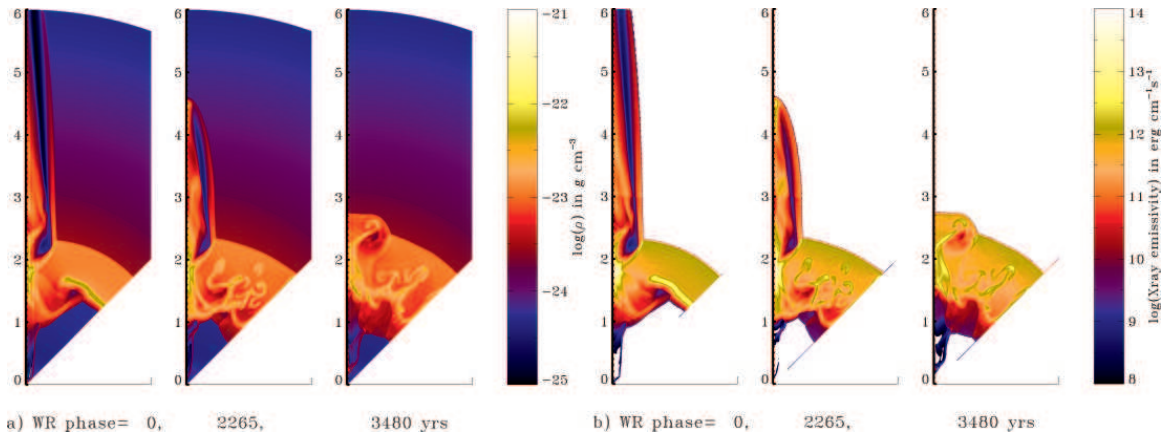}
         \caption{{\em a}) Density and {\em b}) approximate X-ray bremsstrahlung emissivity of the supernova remnant at a time of 330~yr after explosion. The left panels of the density and X-ray figures show the remnant that results from the evolution of asymmetric ejecta in a CSM without a Wolf-Rayet phase, i.e. a pure RSG wind. The middle plots show the case where the Wolf-Rayet phase has lasted 2265 yr, and the right panels show the remnant of an explosion into a CSM with a Wolf-Rayet phase of 3480 yr.  In all of these cases the density of the jet is equal to the density in the rest of the SN, and the velocity is enhanced with a factor 6, giving a jet energy of $5.5 \times 10^{48}$~erg with a maximal velocity of $90$~Mm~s$^{-1}$. The forward shock of the remnant in all three cases is located at a distance of $\sim 2.4$~pc. The reverse shock, however, is located increasingly farther inward for longer WR phases, due to the extra pressure created by collision with the shell. In the left panel the reverse shock is at a distance of $\sim 1.6$~pc, whereas for the middle and right panels it is located at around $1.0$ and $0.8$~pc. The contact discontinuity of the main remnant for the left-most plots is located at a distance of $1.8$~pc, right in front of the high density shell. The contact discontinuity in the middle and right-hand panels is rugged due to the presence of the shell, but is on average located near the high-density fingers. For calculating the X-ray emissivity, the electron temperature is set to be equal to the plasma temperature. Note that the emissivity is not integrated over the line of sight but represents a slice through the meridional plane.
              }
   \label{fig:rhoprofiles}
   \end{figure*}

In Figure~\ref{fig:rhoprofiles}{\em a} the logarithm of the density is plotted in 2D. The three panels show the evolution of the same, axisymmetric, supernova ejecta into different CSMs. The CSMs vary only in the duration of the WR phase. The left plot shows the density after explosion into a plain RSG wind. In the middle and right plots, the progenitor star model did include a WR phase with a duration of respectively 2265 and 3480 years. The average density profiles of the different CSMs were shown in Figure~\ref{fig:WRshell}{\em a}. We find that the jet does not protrude through the shell if the CSM contains a thick, high density, and therefore massive shell. This is the case in the right panel of Figure~\ref{fig:rhoprofiles}{\em a}. The jet can punch through a less massive shell, but it will be shorter than in the case where the CSM was shaped solely by a RSG wind. The forward shock of the main remnant initially is perturbed when it collides with the circumstellar shell, but the shock smoothens within about 100 years due to the high sound speed (in the earliest phase of the remnant $c_s \approx 10^8 - 10^9$~cm~s$^{-1}$).

The thermal emission due to bremsstrahlung is plotted in Figure~\ref{fig:rhoprofiles}{\em b}. Note that the thermal bremsstrahlung emissivity is only an approximation. In reality the X-ray emissivity is determined not only by the temperature and the density, but also by the composition of ejecta and CSM, equilibration of electron- and ion-temperatures, and non-equilibrium ionization, all of which are not taken into account. However, Figure~\ref{fig:rhoprofiles}{\em b} shows that the jet emissivity is small compared to that of the high density parts of the shell, as is indeed what we see in Cas A. In the left and middle panel, most of the thermal emission of the jet is concentrated in the lower first half of the jet. Likely the outer part of the jet would not be bright enough to see in the observations. Therefore, the observations will probably not show the maximal extent of the jet, but rather only a part of it. The shock at the tip of the jet may be seen in non-thermal X-ray emission, but currently there are no detailed X-ray observations of this region. Because of the high temperature near the forward shock, the high density material in the outer part of the remnant stands out compared to the pure-density plot. What appears as a bright shell in the observations and is often cited as the reverse shock may therefore rather be closer to the contact discontinuity, around the higher-density fingers. Tracking of the different fluids would be required to determine exactly where the ejecta and the CSM meet. For a short WR phase (middle and right panels) the clumpiness appears that could be responsible for the observed bright nitrogen-rich knots in Cas A. This is in contrast with what happens for a smooth circumstellar medium shaped solely by a RSG wind, as shown in the left panels. In fact, the contact discontinuity in this case is Rayleigh-Taylor unstable, which, because of limited resolution,
does not show up unless we initially add 1\% density perturbations in the RSG wind.

%-------------------------------------------------------------
%\clearpage
%\clearpage
%
%-------------------------------------------------------------

\subsection{Shell Density and Jet Survival}
\label{sec:SDJS}

In the previous section, we showed that the jet remains present in the SNR only if the progenitor has a very short WR phase, while it is stalled for a progenitor with a longer WR phase. To understand the physics behind this, we compare the energy of the jet with the energy that is required to accelerate the part of the shell within the opening angle of the jet to typical post-shock velocities. The jet does not significantly slow down before it reaches the shell, so for the post-shock velocity we take $3/4$ of the initial jet velocity $v_{\rm j}$. The mass in the shell is calculated by evaluating the density and the volume of each grid cell that lies within the forward and the reverse shock of the shell and within the opening angle of the jet $\theta_{\rm j}$, and correct it for the solid angle. The energy that is needed to accelerate that portion of the shell to $3 v_{\rm j}/4$ should then be:
\begin{eqnarray}
E_\textrm{acc} = \frac{1}{2}M_{\rm shell}(\Omega_{\rm j}) \left(\frac{3}{4} v_{\rm j}\right)^2.
\end{eqnarray}
Here $M_{\rm shell}(\Omega_{\rm j})$ represents the mass of the shell within the solid opening angle of the jet. 

\begin{figure}[!htbp]
   \centering
   \plotone{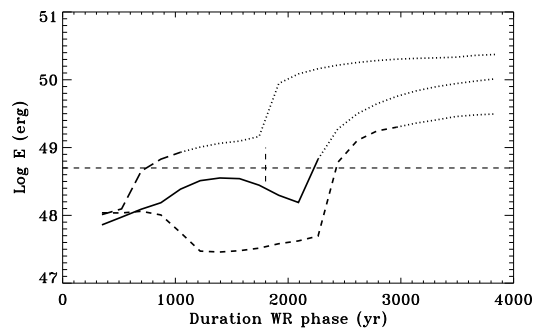}
      \caption{Energy required to accelerate the shell in the jet-region to velocities of $3 v_{\rm jet}/4$, compared to the jet-energy (horizontal dashed line), for different implementations of the progenitor's mass loss history. Curve `b' represents the simulation of the WR shell that was plotted in Figure~\ref{fig:WRshell} and used for the subsequent supernova simulations as plotted in Figure~\ref{fig:rhoprofiles}. Curve `c' represents a simulation of the WR shell without initial density perturbations. Curve `a' represents a simulation of WR shell with different initial density perturbations from the solid-line simulation. The dotted part of the line indicates where the jet starts to be effectively dispersed into the rest of the remnant. Starting from a WR phase of about 1000--3000~years a jet with this energy does not protrude anymore. The vertical patch of line marks the time where analytically the mass in the shell should be high enough to match the energy in the jet.
              }
         \label{fig:energyneeded}
   \end{figure}

The ``required acceleration energy'' is plotted in Figure~\ref{fig:energyneeded} for three simulations of the earliest phase of the WR shell, and compared to the energy in the jet. 
It appears that this gives a good measure for determining whether the jet will remain collimated after the encounter with the shell, or not. When the energy in the jet is higher than the energy that is required to accelerate the portion of the shell, it protrudes, whereas when the acceleration energy needed is much higher, the jet will disperse into the rest of the remnant. It is not a perfect indication; the jet can survive when the energy needed to accelerate the shell is up to a few times higher, although the length of the protruding jet will be limited. 

The dotted portions of the curves indicate the situation where the jet does not protrude, whereas the solid and dashed curves indicate the region where the jet is larger than $\sim 3.8$~pc. In order to rule out a coincidental correlation, we did three different simulations of the shell in the WR phase. We varied the initial density perturbations in the different runs. As a result, the density fluctuations in the shell, and hence the mass accumulation, differ for the three cases. On average, the mass in the shell should be approximately equal to the mass of the RSG wind out to a radius equal to the outer edge of the shell; since the velocity of the WR wind is much higher than that of the RSG wind, the density of the WR wind is lower and it's mass contribution negligable compared to the mass of the RSG that is swept up by it. Therefore, on average, the total energy required to accelerate the shell increases with the time as the WR shell is allowed to develop. The dotted parts of the curves indicate the time where the jet does not have enough energy to break through the shell. The curve that starts as a solid line represents the WR shell simulation that was plotted in Figure~\ref{fig:WRshell} and was used for the SNR/jet simulations that were depicted in Figure~\ref{fig:rhoprofiles}. The horizontal dashed line indicates the energy of the jet used in the simulations. 
The vertical patch of line indicates the analytical solution where the mass, were it to accumulate homogeneously, i.e. without instabilities and equal for each unit angle, multiplied by $\frac{1}{2} (\frac{3}{4}v_{\rm j})^2$ equals the jet-energy. More explicitely, where: 

\begin{eqnarray}
E_{\rm jet} &=& \frac{1}{2} M_{\rm rsg}(R_{\rm shell},\Omega_{\rm j}) \left(\frac{3}{4}v_{\rm j}\right)^2\\
&=& \frac{\Omega_{\rm j} \dot M_{\rm RSG} R_{\rm shell}}{4 \pi v_{\rm RSG}}.
\end{eqnarray}

We can see that for a WR phase of about 2000~years, a jet energy of less than $10^{49}$~erg ($\sim 1.3 \times 10^{53}$~erg isotropic energy), is not sufficient. The exact time varies with each simulation, due to small random differences in the development of the WR shell. However, in all three cases, the turning point lies close to the analytical value of about 1800 yr. The result of the jet-development we have seen in Figure~\ref{fig:rhoprofiles}. Indeed the jet protrudes in the case of a very short WR phase progenitor, and is stalled when the progenitor's WR phase is longer than about 2000~yr.

Note that the critical duration of the WR phase for which a shell forms that is massive enough to block the jet, depends on the exact properties of the progenitor winds. Although in our model the properties of the RSG wind are well constraint by the observations, the properties of the WR wind are more uncertain. The physically important factor is the mass contained in the swept up shell. For a higher mass loss rate or higher velocity of the WR wind, the critical duration of the WR phase may be found to be somewhat shorter. A lower mass loss rate or lower velocity will result in later accumulation of the required mass and induce a longer critical duration of the WR phase.
We simulated jets with energies in excess of $10^{49}$~erg, but find that when the jet energy becomes comparable to the explosion energy, the jet and the SNR shell broaden and no longer resemble the morphology of Cas A. 
   
In the simulations, the assumption of polar symmetry requires us to use a reflective boundary condition on the symmetry axis, which causes material to slowly accumulate in the polar regions. This makes it more difficult for the jet to blast through. However, in our earlier low resolution simulations, we have also simulated the progenitor evolution in the equatorial plane, thus avoiding the boundary condition problem. Although the details are underresolved, the jet still does not survive a massive WR shell. In this case the duration of the WR phase after which the jet stalled was about 5000~yr, which may be an over-estimate due to the low resolution and therefore slower development of the shell.
%-------------------------------------------------------------
%\clearpage
%\clearpage

The duration of the WR phase cannot be determined exactly, because it is dependent on the development of the instabilities. This could explain the difference in the observed morphology of the two jet regions in Cas A (Fig.~\ref{fig:casa_SiMg}). The northeast jet may have encountered a less massive portion of the shell, whereas the southwest jet may have been blocked by a density enhancement in the shell, thus creating a less distinct jet with a more dispersed morphology. In fact, it is found that the velocity of the shock in the west of Cas A is indeed lower than in the east (e.g. \citet{1998Vinketal}).

%\clearpage   
\begin{table*}[!htbp]
\centering                 
\caption{\textrm{Properties of WR shell and forward and reverse shock and the blast wave of the jet after $t=330$~yr and for $E_\textrm{jet}=5.5 \times 10^{48}$~erg.}}            
\label{table:shocks}        
\begin{tabular}{c  c  c  c  c  c  c  c}       
\tableline\tableline               
WR Phase & $R_\textrm{shell}$ & $R_\textrm{forward}$ & $v_\textrm{forward}$ & $R_\textrm{reverse}$ & $v_\textrm{reverse}$ & $R_\textrm{jet}$ &  $v_\textrm{jet}$\\   
 (yr) & (pc) & (pc) & (cm~s$^{-1}$) & (pc) &  (cm~s$^{-1}$) & (pc) & (cm~s$^{-1}$)\\
\tableline
  $0$&  $0$ &  $ 2.43 $&$  5.43 \times 10^8 $&$  1.63  $&$  3.16 \times 10^8  $&$  > 6  $&$1.40 \times 10^9  $\\
  $1569$&  $0.26$ & $ 2.44 $&$  5.34 \times 10^8 $&$  1.28   $&$  2.57 \times 10^8$&$ 4.25$&$  1.03 \times 10^9   $\\
 $1743$&  $0.28$ &  $ 2.45 $&$  5.33 \times 10^8  $&$  1.24 $&$   2.27 \times 10^8 $&$ 5.54 $&$ 1.24 \times 10^9 $\\
 $2091$& $0.33$ &   $ 2.49 $&$ 5.53 \times 10^8  $&$  1.07 $&$ 2.45 \times 10^8  $&$  > 6 $&$ 1.45 \times 10^9  $\\
 $2266$&  $0.35$ &  $ 2.50 $&$ 5.93 \times 10^8 $&$   0.93 $&$ 9.90\times 10^7  $&$  4.57 $&$ 9.90 \times 10^8  $\\
  $2440$&  $0.37$ & $ 2.47 $&$  4.94 \times 10^8 $&$ 0.90$&$ 1.98 \times 10^8   $&$  3.58 $&$ 7.93 \times 10^8    $\\
 $2614$&  $0.40$ &  $ 2.47 $&$  5.24 \times 10^8  $&$ 0.96  $&$ 1.88 \times 10^8 $&$  3.17 $&$ 7.56 \times 10^8   $\\
 $3486$ & $0.50$ &   $ 2.49  $&$  5.33 \times 10^8$&$  0.95   $&$ 1.28 \times 10^8 $&$  2.73 $&$ 5.73 \times 10^8  $\\
\multicolumn{2}{l}{{\em From observations:}} & $ 2.5 $&$ 5.0 \times 10^8 $&$ 1.4 $&$ 3.0 \times 10^8 $&$ > 3.8 $& \ldots \\
\tableline                                   
\end{tabular}
      \tablecomments{Properties of the SNR and the jet depend on the CSM the blastwave runs into. The extent of the WR shell before the explosion is given in the second column. The amount of mass swept up in the shell is essentially the mass of the RSG wind within this radius. The WR wind does not contribute much mass. The bottom row shows the preferred value of the SNR parameter, from observational constraints. The jet puts tighter restrictions on the duration of the WR phase than the properties of the blast wave and reverse shock alone. }
\end{table*}   
%\clearpage

The presence of a circumstellar shell also influences the properties of the blast wave and reverse shock. For a longer WR phase the equatorial blast wave reaches further while the radius of the reverse shock decreases. This is summarized in Table~\ref{table:shocks} for an explosion with jet mass and energy of $3.8 \times 10^{29}$~g respectively $5.5 \times 10^{48}$~erg. Because of the uncertainty in the observational constraints on the properties of the reverse shock, for all the scenarios in Table~\ref{table:shocks} the properties of the forward and reverse shock are within reasonable boundaries. It seems that the presence of a non-relativistic jet constrains the maximal duration of the WR phase more strictly.

\newpage
\section{Conclusions and Discussion}

We have simulated the evolution of axisymmetric ejecta, such as may result from a bipolar supernova explosion, in the context of Cas A and the presence of a jet / counter-jet in this SNR.  For the initial conditions we used a realistic progenitor evolution for a $\sim 20$~M$_\odot$ star, consisting of a RSG wind, followed by a WR wind. 
We find that the presence of a WR shell limits the survival of the jets. The survival depends critically on the energy available in the jet region and on the mass contained in the WR shell. The latter is determined by the duration of the WR phase and the properties of the progenitor winds. For the parameters chosen for the Cas A progenitor, we find that if the WR phase is longer than $2000-5000$~yr, the jets do not protrude through the shell, in which case the situation does not correspond to the presence of jets in Cas A. Therefore, either the progenitor went through a very short WR phase, or it did not have one and exploded as a RSG. In general however, this means that, also if a SNR appears symmetric, the explosion may still have been accompanied by jets.

In order to match the length of the observed jets of Cas A, an energy of at least $2.0 \times 10^{48}$~erg per jet is required. In case a WR shell is formed, higher energies are needed. However, for a WR duration in excess of maximally 5000~yr, the properties of the remnant and the jet do not match the observations. The upper limit we find corresponds to the findings of van Veelen \& Langer (2008, in preparation), where they find that the properties of the forward and reverse shock do not agree with observations for a WR phase that lasts more than 5000~years.

The question now arises if a scenario involving a very short WR phase is realistic. If the progenitor was not in a binary, the chance of having a very short WR phase is very low. The main reason for invoking a WR phase at all is that it explains the clumpiness in the remnant, the lack of hydrogen in the ejecta, the presence of metal clumps far out in the ejecta, and the high N/H ratio in the CSM. A single star at the end of the RSG phase with a clumpy mass loss history \citep{2003ChevalierOishi} may also be able to partly explain the above, but requires equally coincidental circumstances. \citet{2007LamingHwang} favor a short WR phase because, based on the temperature and ionization age of the X-ray emitting gas, they find that the ejecta expanded in a bubble of around $\sim 0.2$~pc., in our case corresponding to a WR phase of about 1000 yrs. They however do not take into account the presence of a shell around the bubble. 

As an alternative to the single star model, a model with a binary companion has been proposed \citep{2006Youngetal}. A common envelope (CE) phase in a close binary solves a number of problems: It explains the low ejecta mass in conjunction with a MS mass of $\sim 20$~M$_\odot$, and provides a natural explanation for a very short  WR phase of the primary star (Podsiadlowski, private communication). However, the details of CE evolution are not well understood and no companion star has currently been found. 

The simulated jets resemble the observed jet of Cas A in opening angle for a variety of parameters. This reinforces the idea that the explosion itself was intrinsically bipolar. 
It remains an interesting question what mechanism is responsible for such an asymmetric explosion, and whether it is related to other bipolar explosion phenomena such as X-ray flashes or LGRBs. Although our case does not resemble the relativistic scenarios as are invoked in models for gamma ray bursts \citep{2005Piran}, it does not seem unreasonable that {once again rotation is involved in creating the asymmetry in a low-energy explosion like this}. Unless the rotation was created during the explosion \citep{2007BlondinMezzacappa}, rotation may have also left an imprint on the CSM and a combination of asymmetric CSM and an asymmetric explosion could have been responsible for the Cas A morphology. 
The observed point source near the center of the remnant is likely to be a neutron star. However, the absence of a bright pulsar wind nebula suggests that the present rotation period of the neutron star is relatively low \citep[$> 160-330$~ms][]{1988SewardWang,2007Vink}. This seems at odds with a rapid rotation of the stellar core as a mechanism to create a bipolar explosion. However, this discrepancy may be solved if the point source is a magnetar that has considerably slowed down.

In conclusion, we would like to emphasize that the presence of jets in Cas A, together with considerable knowledge about nucleosynthesis yields, explosion energy, and compact object, makes this SNR a unique object to investigate the mechanism behind bipolar explosions, and the type of progenitors that produce bipolar explosions.
 
 \acknowledgements
This study has been financially supported by a Vidi grant from the Netherlands Organisation for Scientific Research (NWO). This work was sponsored by the Stichting Nationale Computerfaciliteiten (National Computing Facilities Foundation, NCF) for the use of supercomputer facilities, with financial support from the Nederlandse Organisatie voor Wetenschappelijk Onderzoek (Netherlands Organization for Scientific Research, NWO). We thank Yuri Levin and Bob van Veelen for helpful discussions.

\appendix
\section{Length-Energy relation for a Conical and Cylindrical Jet}
\label{app:lEjet}

The length-energy relation for a jet has been derived in the approximation of a conical and cylindrical morphology for the jet. We use assumptions similar to those that lead to the well-known Sedov-Taylor solution for a spherical blast wave, i.e. a jet interior that is significantly over-pressured with respect to the surrounding medium and no significant radiative losses. For a jet with a constant opening angle, energy conservation gives:

\begin{eqnarray}
&&E=\frac{1}{2}M v^2 + \frac{\Omega}{4\pi} \frac{4\pi R^3}{3} \frac{P}{\gamma -1}.
\end{eqnarray} 
We assume pressure equilibrium between the SNR interior and the shell, and take for the velocity the post-shock velocity, thus getting:
\begin{eqnarray}
E& =&C_\gamma \frac{\Omega}{3} R^3 \rho \left(\frac{dR}{dt}\right)^2\\ 
R^{1/2} dR &=&  \left(\frac{3 E}{ C_\gamma \Omega A}\right)^{1/2} dt ,\qquad \Omega \approx \pi \theta_{\rm j}^2\\
R&\approx&\left( \frac{27 E t^2}{4 C_\gamma \pi \theta{\rm j}^2 A} \right)^{1/3}\\
R&\approx&1.15\left( \frac{E t^2}{ \theta_{\rm j}^2 A} \right)^{1/3},
\end{eqnarray}
Here $\theta_{\rm j}$ is the jet's opening angle, $E$ the jet-energy, $R$ the length of the jet at time $t$, $r$ the radius of the cone, $\rho = A/R^2$ with $A$ a normalization constant, and $C_\gamma = (4 \gamma)/((\gamma-1)(\gamma+1)^2) = 45/32$ for a polytropic index of $\gamma=5/3$.

For a cylinder we use the same approach, but because of the different volume we get a different relation between the length and energy:

\begin{eqnarray}
E=C_\gamma M \left(\frac{dR}{dt}\right)^2.
\end{eqnarray}
The mass of a cylinder-like jet is:
\begin{eqnarray}
M&=&\int^{R_1}_{R_0} \pi r^2 \rho_0 dR \\
&=&\int^{R_1}_{R_0} \pi r^2 \frac{A}{R^2} dR \\
&=&\pi r^2 A \left(\frac{1}{R_0}-\frac{1}{R_1}\right).
\end{eqnarray}
Energy conservation is then given by:
\begin{eqnarray}
E & = & C_\gamma A \pi r^2\frac{R-R_0}{R R_0} \left(\frac{dR}{dt}\right)^2 \qquad \frac{E}{C_\gamma A \pi r^2} = \epsilon \\
\epsilon  & = & \frac{R- R_0}{R R_0 }\left(\frac{dR}{dt}\right)^2.
\end{eqnarray}
Substitution of $x=\frac{R}{R_0}$ with changing the boundaries from $(R_0,R)$ accordingly to $(1,\frac{R}{R_0})$ leads to:
\begin{eqnarray}
\epsilon & = & \frac{x- 1}{x} R_0 \left(\frac{dx}{dt}\right)^2  \qquad \tilde \epsilon = \frac{\epsilon}{R_0} \\
\sqrt{\tilde \epsilon} & = & \sqrt{\frac{x- 1}{x}} \frac{dx}{dt} .
\end{eqnarray}
This can be reduced to quadrature by substituting $x=y^2$:
\begin{eqnarray}
\int^{\sqrt{R/R_0}}_1 \sqrt{y^2-1} dy & = & \frac{1}{2}\sqrt{\tilde \epsilon}\  t .
\end{eqnarray}
The left side is a standard integral:
\begin{eqnarray}
\int^{\sqrt{R/R_0}}_1 \sqrt{y^2-1} dy & = & 
   \left. \frac{1}{2} y \sqrt{y^2-1}-\frac{1}{2}\ln\left(y+\sqrt{y^2-1}\right) \right]^{\sqrt{R/R_0}}_1.
\end{eqnarray}
Thus the solution becomes:
\begin{eqnarray} 
\left.y \sqrt{y^2-1}-\ln\left(y+\sqrt{y^2-1}\right) \right]^{\sqrt{R/R_0}}_1 = \sqrt{\frac{E t^2}{C_\gamma A \pi r^2 R_0}}.
\end{eqnarray}
Restoring the original variables gives:
\begin{eqnarray} 
\left.\sqrt{\frac{R}{R_0}} \sqrt{\frac{R}{R_0}-1}-\ln\left(\sqrt{\frac{R}{R_0}}+ \sqrt{\frac{R}{R_0}-1}\right) \right]^R_{R_0} = \sqrt{\frac{E t^2}{C_\gamma A \pi r^2 R_0}},
\end{eqnarray}
For $R \gg R_0$ this can be approximated by:
\begin{eqnarray} 
\frac{R}{R_0}-\ln[2 \sqrt{\frac{R}{R_0}}] = \sqrt{\frac{E t^2}{C_\gamma A \pi r^2 R_0}},\\
R =  \sqrt{\frac{E t^2 R_0}{C_\gamma A \pi r^2}} + R_0 \ln \left(2 \sqrt{\frac{R}{R_0}}\right).
\end{eqnarray}

The second term on the right-hand side has only a very weak dependence on $R$ and in the limit $R \gg R_0$ it is negligable compared to the first term on the right-hand side.

\bibliography{adssample}
%------------------------
%------------------------

\clearpage

\end{document}